\documentclass{article}
\usepackage{spconf,amsmath,graphicx}
\usepackage{cite}
\usepackage{algorithm,algcompatible}
\usepackage{graphicx}
\usepackage{textcomp}
\usepackage{xcolor}
\usepackage{enumitem}
\usepackage{amsmath,amssymb}
\usepackage{float}
\newfloat{algorithm}{t}{lop}
\newcommand{\norm}[1]{\left\lVert#1\right\rVert}
\newtheorem{theorem}{Theorem}[section]

\renewcommand{\COMMENT}[2][.3\linewidth]{%
  \leavevmode\hfill\makebox[#1][l]{\#~#2}}

\title{Optimal Transport Based Change Point Detection and Time Series Segment Clustering}
\name{Kevin C. Cheng$^{\star}$, Shuchin Aeron$^{\star}$, Michael C. Hughes$^{\star}$, Erika Hussey$^{\dagger}$,  Eric L. Miller$^{\star}$ \thanks{This research was supported by funding from Army Research Center Natick via Tufts Center for Applied Brain and Cognitive Sciences (CABCS) under ARM994. Shuchin Aeron was supported in part by NSF CAREER award.}}
\address{$^{\star}$Tufts University, $^{\dagger}$CCDC-Soldier Center \\
    $^{\star}$$^{\dagger}$\{\textit{first name}\}.\{\textit{last name}\} @ tufts.edu}
\begin{document}
\maketitle

\setlength{\abovedisplayskip}{2pt plus 3pt}
\setlength{\belowdisplayskip}{2pt plus 3pt}

\begin{abstract}
Two common problems in time series analysis are the decomposition of the data stream into disjoint segments that are each in some sense ``homogeneous'' - a problem known as Change Point Detection (CPD) - and the grouping of similar nonadjacent segments, a problem that we call Time Series Segment Clustering (TSSC). Building upon recent theoretical advances characterizing the limiting \emph{distribution-free} behavior of the Wasserstein two-sample test (Ramdas et al. 2015), we propose a novel algorithm for unsupervised, distribution-free CPD which is amenable to both offline and online settings. We also introduce a method to mitigate false positives in CPD and address TSSC by using the Wasserstein distance between the detected segments to build an affinity matrix to which we apply spectral clustering. Results on both synthetic and real data sets show the benefits of the approach. 

\end{abstract}


\section{Introduction}
Change point detection (CPD) is a fundamental problem in data analysis with implications in many real world applications including the analysis of financial\cite{carvalho_simulation-based_2007}, electrocardiogram (ECG) \cite{UCRArchive2018}, and human activity data \cite{li_m-statistic_2015}. Given a collection of change points, time series segment clustering (TSSC) seeks to group nonadjacent periods of activity which are, in some sense, ''similar,'' in an unsupervised manner. Applications here overlap with those of CPD 
\cite{aghabozorgi_time-series_2015}. 

In this paper, we focus on the use of statistical methods for CPD which are broadly classified as either parametric (model-based) or non-parametric \cite{wald_sequential_1947,basseville:hal-00008518}.  The  problem formulation employed by the majority of methods takes the observations as a sequence of random variables whose distribution changes abruptly at unknown points in time.  The processing goal for CPD is to determine when the switches occur and, in those instances where TSSC is required, use a similarity measure to cluster like segments. 

Parametric methods employ a specific model for the dynamics of the time series (either assumed \cite{chamroukhi_joint_2013} or learned from data \cite{lee_time_2018}) and then make use of decision theory to identify change points.  Classically, ARMA-type models and their state-space generalizations were the basis for parametric efforts starting in \cite{willsky_generalized_1976} with recent work focusing on hierarchical models such as switching linear-dynamical systems (SLDS) \cite{murphy_switching_1998}. Generally, parametric methods are effective when the modelling assumptions hold. For example, SLDS assumes geometric state duration distributions and Gaussian observation models. When these assumptions are not applicable, performance will likely suffer.

When the dynamics or observations cannot be easily modeled, we can consider distribution-free methods that do not assume any particular parametric family of distributions. Change points are then estimated from sample distributions using density-ratio estimates \cite{liu_change-point_2013, kanamori_least-squares_nodate} or through two-sample tests like maximum mean discrepancy (MMD) \cite{gretton_kernel_nodate}, which was recently used for non-parametric CPD~\cite{li_m-statistic_2015}.

Similar to CPD, parametric TSSC methods have been explored using ARMA based models \cite{corduas_time_2008} or HMMs \cite{oates_clustering_1999}. Non-parametric TSSC methods generally use alternate representations of time series such as frequency-based wavelet decompositions \cite{huhtala_mining_1999} or distribution-based methods \cite{dahlhaus_kullback-leibler_1996}.

In this work, we contribute a new set of non-parametric CPD and TSSC methods based on recent statistical results in the theory of Optimal Transport (OT).  Assuming independent and identically distributed (IID) data, Ramdas et al. \cite{ramdas_wasserstein_2015} provides a theoretical analysis of the asymptotic distribution of an OT-based two-sample test under the null hypothesis for deciding whether two empirical probability density functions are from the same distribution. We use this result as the basis for a sliding window test for identifying change points in a scalar time series. Another novel aspect of our method is the development of a statistically-derived ''matched filter'' for post-processing  our OT statistic to reduce false positives. Given the identified change points, we develop an OT-based spectral clustering scheme for TSSC.

\begin{figure*}[]
\centering
\includegraphics[width=2.0\columnwidth]{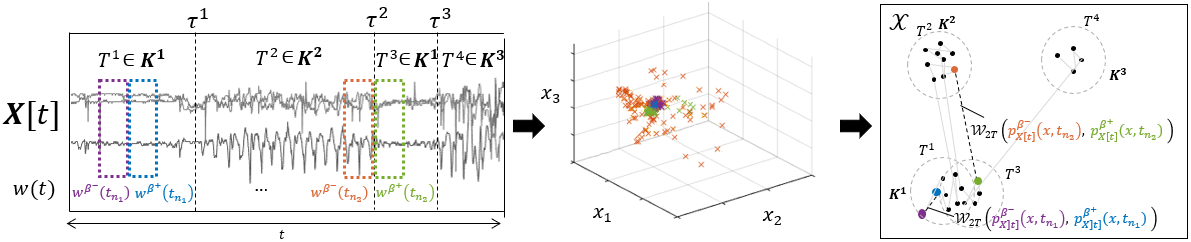}
\caption{The time series $X[t] \in \mathbb{R}^3$ can be decomposed at change points $\tau^i$ into 4 segments $T^i$ represented by 3 actions. Sample windows of time series data (left) are represented as point cloud in $\mathbb{R}^3$ (center) which in turn corresponds to a single point in the space of all probability measures $\mathcal{X}$ (right), which is an estimate of the process distribution. The proposed method uses the Wasserstein two-sample test between adjacent windows on each dimension independently as the change point test statistic. At $t_{n_1}$ the two windows belong to similar distributions and thus no change is detected. However at $t_{n_2}$ spanning the change point $\tau^2$ places the distributions of adjacent windows in different clusters $K^2, K^1$ thus resulting in a high CPD statistic. }
\label{fig:DataModel}
\end{figure*}

To organize this paper, we start with an overview of optimal transport concepts followed by problem formulation for CPD and TSSC. We then detail our proposed method and evaluate our techniques on toy and real-world data sets. We show improved precision and recall for CPD (as summarized in F1 scores) compared to state of the art. We also show improved label accuracy in TSSC for human activity data.

\section{Optimal Transport Background}  \label{sec:OT}
Given two probability distributions $p(x), q(y)$, where $x,y \in \mathbb{R}^d$, the 2-Wasserstein distance, or earth mover's distance, $\mathcal{W}_2(p(x),q(y))$ is defined as the minimum expected squared Euclidean cost required to transport $p(x)$ to $q(y)$. Formally,
\begin{align} \label{eq:emd}
\mathcal{W}_2(p(x),q(y)) = \min_{\pi \epsilon \Pi} \int_x \int_y \norm{x-y}_2^2\pi(x,y) dx dy, \nonumber \\
\int_y \pi(x,y)=p(x), \,\,  \int_x \pi(x,y) =q(y)  
\end{align} 
where $\Pi$ denotes the set of all joint distributions. It is well-known that \eqref{eq:emd} is a linear program. Further, $\mathcal{W}_2(\cdot,\cdot)$ is a metric on the set of probability distributions \cite{peyre_computational_2018} and metrizes weak convergence of probability measures.

We employ a distribution-free, non-parametric Wasserestein two-sample test (W2T) as a discrepancy measure  between two sets of points. To this end, we note the following:
\begin{theorem}\label{eq:2Samp} (From \cite{ramdas_wasserstein_2015})
Under the null hypothesis $H_0: P=Q$, given empirical CDF's $P_m$, $Q_n$ consisting of $m, n$ IID samples from scalar distributions $P$, $Q$, $\mathcal{W}_{2T}(P_m,Q_n)  = \frac{mn}{m+n} \int_0^1 (P_m(Q_n^{-1}(x))-x)^2 dx  \overset{d}{\longrightarrow} \int_0^1 \mathcal{B}^2(x)dx = \mathbb{B}_2$
\end{theorem}
where $\mathcal{B}(x)$ denotes the standard Brownian motion. From  \cite{tolmatz_distribution_2002}, $\mathbb{B}_2$ has mean $\mu_{\mathbb{B}_2} = 0.166$ and that we reject the null with confidence $\alpha=0.05$ using a threshold of $\lambda=0.462$. 

\section{Problem Formulation} \label{sec:problem}
As detailed in Fig.~\ref{fig:DataModel} and throughout, we consider a time series $X[t] \in \mathbb{R}^d$, $t=1,2, \dots$, where the data consists of distinct time segments $[0,\tau^1], [\tau^1+1, \tau^2], \dots, [\tau^{S-1}+1 , \tau^S]$, with $\tau^1 <\tau^2 <\dots$ such that within each time segment, $X[t], t \in [\tau^{i-1}+1, \tau^{i}], i = 1,2,...,S$, are IID samples from one of $k=1, 2, \dots K$ unknown distributions, where we assume here that $K$ is known\textit{ a priori}. The problem of change point detection (CPD) is to estimate $\tau^i$, and the problem of  time series segment clustering (TSSC) is to cluster the $S$ segments into $K$ classes.

\section{Proposed Method}

\subsection{Change Point Detection}\label{sec:CPD}

\begin{figure*}[]
\centering
\includegraphics[width=2.0\columnwidth]{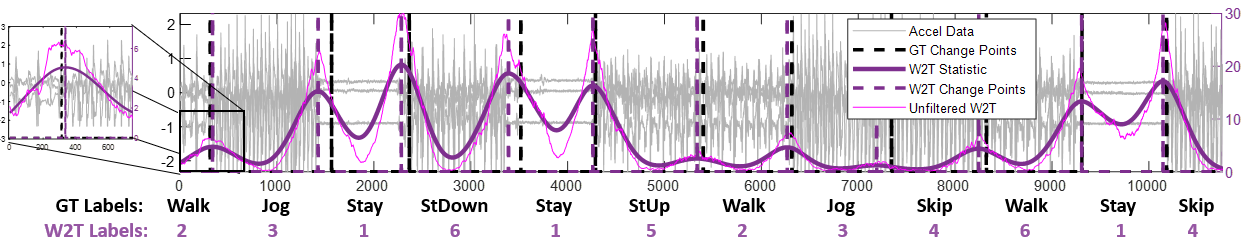}
\caption{CPD and TSSC results from HASC2016-PAC data (black, left axis). For CPD, we plot both unfiltered (thin purple) and match filtered (thick purple) change point statistics (right axis). The left subplot shows how the matched filter removes false positives and improves localization of the change point. For TSSC, the bottom row shows our method's assigned cluster labels, which make only one mistake relative to ground truth (GT) by grouping the stair-down segment with the last walk segment.} 
\label{fig:HascOut}
\end{figure*}

Given time-series $X[t]$, we define two empirical probability density functions (PDFs) at each time $t$ generated from the sum of dirac-delta functions supported on a window of $\beta$ samples collected before and after $t$ yielding $p^{\beta^{\pm}}_{X[t]}(x) = \frac{1}{\beta}\Sigma_{\tau=1}^{\beta}\delta(x-X[t \pm \tau])$. After transforming each PDF $p$ into a cumulative distribution function (CDF) $P$, we can compute a change point statistic from the Wasserstein two-sample test (W2T) between the CDFs of the two windows:
\begin{equation}
    \sigma[t] = \mathcal{W}_{2T} \left(P^{\beta^-}_{X[t]}(x), P^{\beta^+}_{X[t]}(x)\right).
\end{equation}
The nominal, offline approach to CPD is to label local maxima of $\sigma[t]$ that exceed some threshold parameter as change points \cite{li_m-statistic_2015}. Shown through empirical analysis on both simulated and real data, we find this is problematic. Fig.~\ref{fig:HascOut} indicates the presence of spurious local maxima leading to a large number of false alarms and ambiguity in the change point locations. Moreover, the sliding-window nature of the processing causes a change point at time $t$ to create an extended signature in $\sigma[t]$ over the interval $[t-\beta, t+\beta]$.

\begin{figure}
\centering
\includegraphics[width=0.8\columnwidth]{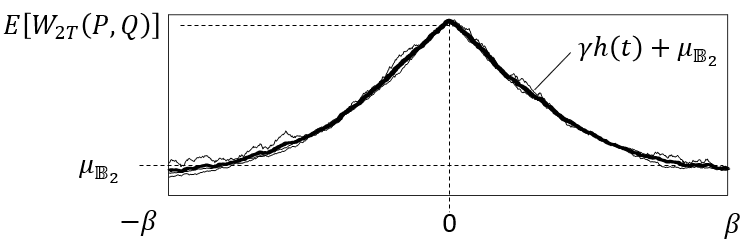}
\caption{Un-normalized empirically estimated matched filter. Given a change point at $t=0$ and window size $\beta$, the effects of the change point are reflected in the W2T statistic on the interval $[-\beta, \beta]$. Thin traces represent the ensemble average of 200 IID sequences with different simulated change points ($N(0,1) \rightarrow N(0.2,1)$, $N(0,1) \rightarrow N(0,1.2)$ and $N(0,1) \rightarrow L(0,\frac{1}{\sqrt{2}})$). The matched filter $h[t]$ is normalized by removing the bias $\mu_{\mathbb{B}_2}$ and scaling by $\gamma$ to have unit area. }
\label{fig:CorrelationFilter}
\end{figure}

These observations suggest the benefit of a matched filtering approach to reduce the spurious maxima and better localize true changes.  In Fig.~\ref{fig:CorrelationFilter} we estimate the shape of this signature empirically by averaging over ensembles of simulated IID data with a known change points separating samples from different pairs of distribution. From this plot, we observe that the structure of this function across a number of distributional changes is remarkably consistent. The theoretical analysis and discussion of this filter is left to future work. We derive the filter $h[t]$ by removing the bias and normalizing by a constant such that the peaks of $\sigma[t]$ signal are preserved. Change points are the set of local maxima where $\sigma[t] \circledast  h[t]$\footnote{Where $\circledast$ denotes the convolution operation} exceeds a threshold: $\{\tau\}=\{t \, | \, peaks( \sigma[t]\circledast h[t]) > \lambda\}$\footnote{ $t \in \{peaks(f[t])\} \, \iff \, f[t]>f[t-1] \And f[t]>f[t+1]$}. 

\subsection{Time Series Segment Clustering}
Given change points $\{\tau^i\}$ and time segments $T^{i}=\{X[t]~|~\tau^i< t \leq \tau^{i+1}\}$, the process distribution in this time segment is estimated by $p^{T^i} = \left(\Sigma_j w_j\right)^{-1}\Sigma_{j=\tau^i}^{\tau^{i+1}}w_j \- \delta(x-X[j])$. This represents a weighted point cloud generated from the data points over the time interval. Samples are weighted by a windowing function that down-weights samples around transition boundaries, mitigating the effects of segmentation errors and non-instantaneous transitions. To this effect, we use a half Hamming window of length $2\beta$ for samples within $\beta$ of either boundary. Samples outside this range have weights $w_j=1$. 

The similarity matrix between time segments $A[i,j] = \exp\left(-\mathcal{W}_2\left(p^{T_i}, p^{T_j}\right)\right)$, uses the 2-Wasserstein distance between their respective empirical distributions as the distance measure. Given the number of action clusters $K$, we utilize the similarity graph structure under the Wasserstein metric by clustering time segments via spectral clustering~\cite{vonluxburg_tutorialSpectralClustering_2007} into the optimal action clusters.

\begin{algorithm}
\caption{Wasserstein Change Point Detection and Time Series Segment Clustering}
\begin{algorithmic}\label{alg:algorithm}
\STATE \textbf{Input}: $X[t]$, $\beta$, $K$, $h[t]$, $\lambda$, $\left\{w_j\right\}_{j=1}^{\beta}$
\STATE \textbf{Output}: $\{\tau\}$, $c$
\FORALL{t}\COMMENT{\textit{CPD}}
\STATE $p_{X[t]}^{\beta^-}(x) = \frac{1}{\beta}\sum_{i=1}^\beta \delta(x-X[t-i])$
\STATE $p_{X[t]}^{\beta^+}(x) = \frac{1}{\beta}\sum_{i=1}^\beta \delta(x-X[t+i])$
\STATE $\sigma[t] = \mathcal{W}_{2T}(P_{X[t]}^{\beta^-}, P_{X[t]}^{\beta^+})$
\ENDFOR
\STATE $\{\mathbf{\tau}\} = \{t\, |\, peaks( \sigma[t] \circledast h[t] ) > \lambda\}$
\STATE $p^{T^i} = \frac{1}{\Sigma_j w_j}\Sigma_{j=\tau^i}^{\tau^{i+1}}w_j\delta(x-X[j])$\COMMENT{\textit{TSSC}} 
\FORALL{$0\leq i,j < |\mathbf{\tau}|$}
\STATE $A(i,j) = exp(-\mathcal{W}_2 (p^{T^i},p^{T^j}))$
\ENDFOR
\STATE $c = SpectralClustering(A, K)$
\end{algorithmic}
\end{algorithm}

\section{Evaluation}

\begin{table*}
\centering
\begin{tabular}{l|ccc||cc||cc||c|cc}
\hline 
    &&&& \multicolumn{2}{c||}{CP-AUC} & \multicolumn{2}{c||}{CP-F1} & \multicolumn{3}{c}{Label Acc}\\

    Data & $K$ & $\beta$ & $\delta$ 
    & W2T & MStat 
    & W2T & MStat 
    & GT
    & W2T & MStat \\
    \hline
    Beedance & 3 & 14 & 14 
        & 0.527 & 0.549 
        & 0.647 & 0.625 
        & 0.705
        & 0.651 & 0.646  
        \\
    HASC-PAC2016 & 6 & 500 & 250 
        & 0.689  & 0.658 
        & 0.748 & 0.713 
        & 0.789
        & 0.658 & 0.675 \\
    HASC2011 & 6 & 500 & 250 
        & 0.576 & 0.585 
        & 0.824 & 0.770
        & 0.565
        & 0.498 & 0.382 \\
    ECG200 & 2 & 100 & 50 
        & 0.585 & 0.584 
        & 0.637 & 0.582 
        & 0.864
        & 0.708 & 0.716 \\

    \hline
\end{tabular}
\caption{CPD evaluation using AUC and F1 for proposed W2T method and MStat for given number of labels $K$, window size $\beta$, and detection delay $\delta$. TSSC is evaluated with label Hamming accuracy using ground truth, W2T, and MStat change points. }
\label{tab:metrics}
\end{table*}

\subsection{Evaluation Criteria}


We use the area under the ROC curve (CP-AUC) to evaluate change point performance, following previous work~\cite{li_m-statistic_2015,chang_kernel_2019,liu_change-point_2013}. We also report the F1 score (CP-F1) for offline multiple CPD  \cite{truong_selective_2018} using a margin of error $\delta$ for the acceptable offset to the true label.

For TSSC, cluster labels are mapped onto the ground truth labels using the standard Munkres algorithm and evaluated using the Hamming distance. Performance is reported in Tab.~\ref{tab:metrics} separately using ground truth change points (GT) and learned change points (W2T or MStat).

\subsection{Experimental Setup}
We compare the performance of our algorithm to the M-Statistic (MStat)\cite{li_m-statistic_2015}, setting parameters $N=1$, $M=\beta$. For fair comparison, we employ a MStat matched filter $h_M[t]$ using a method analogous to that outlined in Sec.~\ref{sec:CPD}. The only hyperparameters to the CPD model are the window size $\beta$ and detection threshold $\lambda$. Since the window size controls the width of the matched filter, we utilize domain knowledge to set $\beta$ based on the expected frequency of changes. We also set the threshold parameter $\lambda=0$ as the distribution of the MStat under the null is not known. The hyperparameters, along with the true positive detection window $\delta$ used for F1 can be found in Tab.~\ref{tab:metrics}. For vectored time series we computed the W2T over each dimension and averaged the result. We evaluate on the following datasets:

\textbf{HASC-PAC2016}: \cite{ichino_hasc-pac2016:_2016} consists of over 700 three-axis accelerometer sequences of subjects performing six actions: 'stay', 'walk', 'jog', 'skip', 'stairs up', and 'stairs down'. We evaluate on the 92 longest sequences.

\textbf{HASC-2011}: three-axis accelerometer data from 6 actions: 'stay', 'walk', 'escalator up', 'elevator up', 'stairs up', and 'stairs down'. 

\textbf{Beedance}: \cite{oh_learning_2008} movements of dancing honeybees who communicate through three actions: "turn left", "turn right" and "waggle". We use the gradient of the data as our input.

\textbf{ECG200}: \cite{UCRArchive2018} detection of abnormal heartbeats in ECG. 

\subsection{Results}
\begin{figure}[]
\centering
\includegraphics[width=0.82\columnwidth]{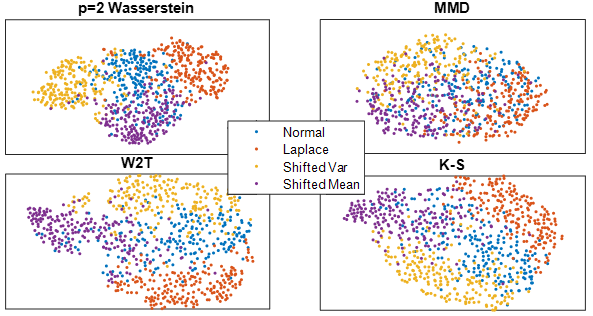}
\caption{t-SNE embedding of simulated data of 200 windows of IID 100 samples from 4 simulated distributions: $\mathcal{N}(0,1)$ (blue), $\mathcal{L}(0,\frac{1}{\sqrt{2}})$ (orange), $\mathcal{N}(0.2,1)$ (yellow), $\mathcal{N}(0,1.2)$ (purple) using Wasserstein metric, and two-sample tests: MMD, W2T, and Kolmogorov-Smirnov} 
\label{fig:SimulatedEmbedding}
\end{figure}
The proposed algorithm demonstrates robust results for CPD and TSSC. Fig.~\ref{fig:HascOut} shows clear detection of change points on HASC-PAC2016, strong efficacy of the matched filter in reducing false positives, and a single label mis-classification.

The CPD performance for the W2T and MStat are comparable under the AUC metric, however, under the F1 metric, W2T consistently performs better. We note that the computation complexity of the W2T ($O(\beta \log(\beta))$) is an improvement compared to that of the MStat ($O(\beta^2)$) and that the OT measures show tighter clustering in the low-dimensional embedding of various simulated measures (Fig.~\ref{fig:SimulatedEmbedding}).

Comparing to results reported in \cite{chang_kernel_2019}, our unsupervised method shows competitive results with an AUC of 0.527 on Beedance compared to supervised parametric models such as ARMA (0.537) and ARGP (0.583). We observe that since $h[t]$ smooths the test statistic, its inclusion decreases AUC for a better F1 score, which we see as a positive tradeoff. For example, when including $h[t]$ for HASC2011, our AUC drops from 0.630 to 0.576 while F1 improves from 0.720 to 0.824. 

In terms of TSSC, using our unsupervised, distribution-free approach, we are able to achieve a 65\% label accuracy on the Beedance data. For comparison, a state of the art supervised parametric model \cite{oh_learning_2008} achieves an 87.7\% label accuracy, and a parametric unsupervised model using switching vector autoregressive HMMs \cite{fox_nonparametric_2009} achieves a label accuracy of 66.8\%. HASC also shows strong performance given that a total of six possible assignments were available.

\section{Discussion}
We propose a distribution-free, unsupervised approach to CPD and TSSC for time-series data. In our experiments, we run the CPD in an offline manner. Applied in an online setting, the minimum detection delay would be $2\beta$.

We approach CPD and TSSC with a weak set of assumptions: that change points occur when the process distribution changes, and actions can be clustered based on their respective empirical distributions. However, clearly time series data is rarely IID. In future work, we will expand these methods for CPD and TSSC beyond IID assumptions.

\bibliographystyle{./bibliography/IEEEbib}
\bibliography{./bibliography/IEEEabrv,./bibliography/IEEEexample}

\appendix{}

\end{document}